\documentclass[11pt,preprint]{emulateapj}
\usepackage[latin1]{inputenc}
\usepackage{amsmath}
\usepackage{amsfonts}
\usepackage{amssymb}
\usepackage{graphics,graphicx}
\usepackage{units}

\begin{document}
\title{Core Gas Sloshing in Abell 1644}
\author{Ryan E. Johnson\altaffilmark{1,2}, 
Maxim Markevitch\altaffilmark{2}, 
Gary A. Wegner\altaffilmark{1},
Christine Jones\altaffilmark{2} and
William R. Forman\altaffilmark{2}
}
\altaffiltext{1}{Department of Physics and Astronomy, Wilder Lab, Dartmouth College, Hanover, NH, 03755}
\altaffiltext{2}{Smithsonian Astrophysical Observatory, Harvard-Smithsonian Center for Astrophysics, Cambridge, MA 02138}
\begin{abstract}
We present an analysis of a 72 ks Chandra observation of the double cluster Abell 1644 (z=0.047).  The X-ray temperatures indicate the masses are M$_{500}=2.6\pm{0.4}\times10^{14} h^{-1} M_{\odot}$ for the northern subcluster and M$_{500}=3.1\pm{0.4}\times10^{14} h^{-1} M_{\odot}$ for the southern, main cluster.  We identify a sharp edge in the radial X-ray surface brightness of the main cluster, which we find to be a cold front, with a jump in temperature  of a factor of $\sim$3. This edge possesses a spiral morphology characteristic of core gas sloshing around the cluster potential minimum. We present observational evidence, supported by hydrodynamic simulations, that the northern subcluster is the object which initiated the core gas sloshing in the main cluster at least 700 Myr ago.  We discuss reheating of the main cluster's core gas via two mechanisms brought about by the sloshing gas: first, the release of gravitational potential energy gained by the core's displacement from the potential minimum, and second, a dredging inwards of the outer, higher entropy cluster gas along finger-shaped streams.  We find the available gravitational potential energy is small compared to the energy released by the cooling gas in the core. 
\end {abstract}

\section{INTRODUCTION}
Double and multiple cluster systems arise as a natural consequence of hierarchical structure formation. Being acutely sensitive to the gas density in clusters, X-ray observations provide a powerful method for unambiguously identifying cluster systems and substructure within them. With early X-ray telescopes such as Einstein, ROSAT and ASCA, $60\%$ of clusters were categorized as ``relaxed'', based on the smooth, mostly circular flux isophotes centered on their cores (e.g. Jones \& Forman 1999). It is only with the most recent generation of X-ray telescopes (\textit{XMM-Newton} and, to a greater degree, \textit{Chandra}) that the angular resolution has increased sufficiently to examine small scale features near the cluster cores. Many clusters previously considered relaxed, upon closer inspection, reveal highly disturbed cores.
The disturbances include nearly ubiquitous AGN-blown bubbles in the intracluster gas (c.f. Churazov et al. 2000, Fabian et al. 2000, Nulsen et al. 2002, and review in McNamara \& Nulsen 2007) and cold fronts (review by Markevitch \& Vikhlinin 2007, hereafter MV07). Markevitch et al. (2003) found that more than 2/3 of the clusters classified as cool core systems in a flux limited sample by Peres et al. (1998) exhibit edges in the X-ray surface brightness profiles around their cores when viewed with \textit{Chandra}. X-ray spectra show that most of these edges are contact discontinuities, or ``cold fronts'' (MV07), though occasionally they are caused by shock fronts (e.g. Hydra A, in Nulsen et al. 2005).

Cold fronts indicate gas motion in a cluster and may be located at the boundary of an infalling subcluster, as in the merger in A3667 (Vikhlinin et al. 2001) or in 1E0657-56 (Markevitch et al. 2002). They may also arise in the core of the primary cluster during the passage of a subcluster, when the core gas starts ``sloshing'' in response to the gravitational disturbance (Markevitch et al. 2001; Churazov et al. 2003; Ascasibar \& Markevitch 2006, hereafter AM06). We find direct evidence in support of this gas sloshing in the core of Abell~1644, which we explore in this paper.

\section {CORE GAS OSCILLATION (SLOSHING)}
``Sloshing'' of the dense cluster core gas was first proposed by Markevitch et al. (2001) to explain a cold front observed in the core of a relaxed cluster, Abell 1795. This process was studied via hydrodynamic simulations in AM06 and reviewed in detail in MV07. A similar process to explain these features, though slightly different in interpretation (MV07), was proposed by Tittley \& Henricksen (2005).

From the above works, it was found that core sloshing in otherwise undisturbed clusters arises when the cluster undergoes a perturbation to its gravitational potential caused by another infalling group or cluster, even if that infalling cluster is gasless and the resulting large scale disturbance is small.  The sloshing occurs as a result of the gas core lagging behind the cluster the potential minimum, as they both move towards the perturbing object.  As the gas core falls back onto the potential minimum, it overshoots it and begins to oscillate.  With each oscillation, the gas core is moving against its own trailing gas, producing an ``edge'' in the X-ray brightness which expands out from the cluster. This sequence of events is described in more detail in AM06.  

The continued oscillation of the core gas about the potential minimum produces a succession of radially propagating cold fronts, manifested as concentric edges in the surface brightness distribution. These fronts may form a spiral structure when the sloshing direction is near the plane of the sky and the merger has a non-zero angular momentum (AM06). A recent study by Lagana et al. (2009) examining substructure in a sample of X-ray clusters has found such a spiral pattern around the cores of those clusters.  Here, we see exactly this spiral structure in the core of the main cluster in Abell 1644.

In $\S$ 3, we describe the reduction of the Chandra data on A1644 and extraction of X-ray spectra. In $\S$ 4 we list the various gas quantities from the fitted models. In $\S$ 5 we discuss implications of these measurements, and our conclusion is in $\S$ 6.

\begin{figure*}[h]
\centering
\includegraphics[scale=0.9]{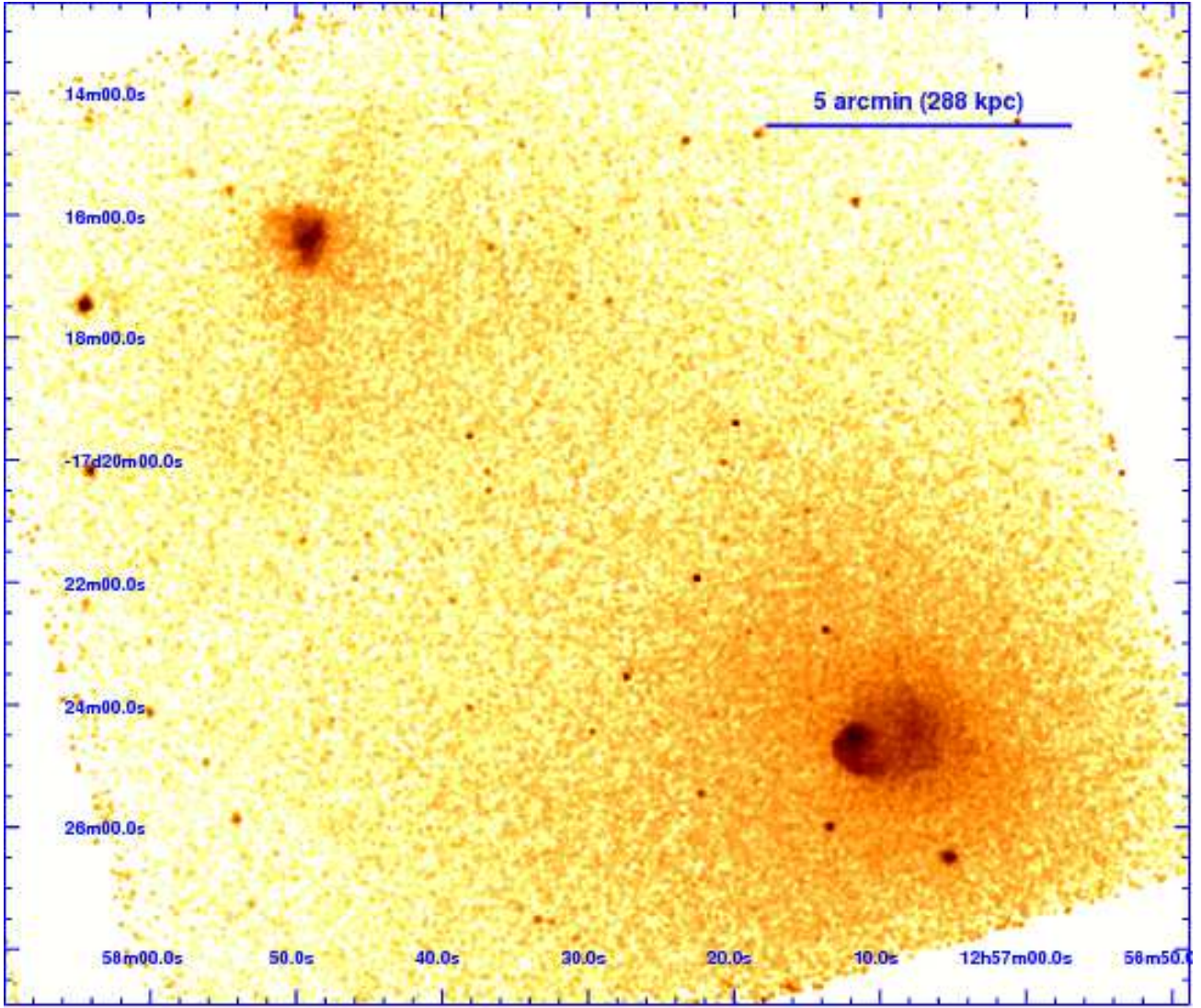}
\caption{\footnotesize{Combined 70ks Chandra ACIS-I image (obsid 2206 and 7922) over the energy range 0.5-2.5 keV. We see the two bright cores of A1644-N and A1644-S. Both the main and subcluster's emission is preferentially extended towards the other, indicating previous interaction between the two systems. We also see the spiral pattern of the gas sloshing in A1644-S. The image has been background subtracted, exposure corrected, binned by a factor of 2 and smoothed in ds9 using a Gaussian kernel with radius 4 pixel ($\sim4\arcsec$).}}
\end{figure*}

\begin{figure*}[h]
\includegraphics[scale=0.85]{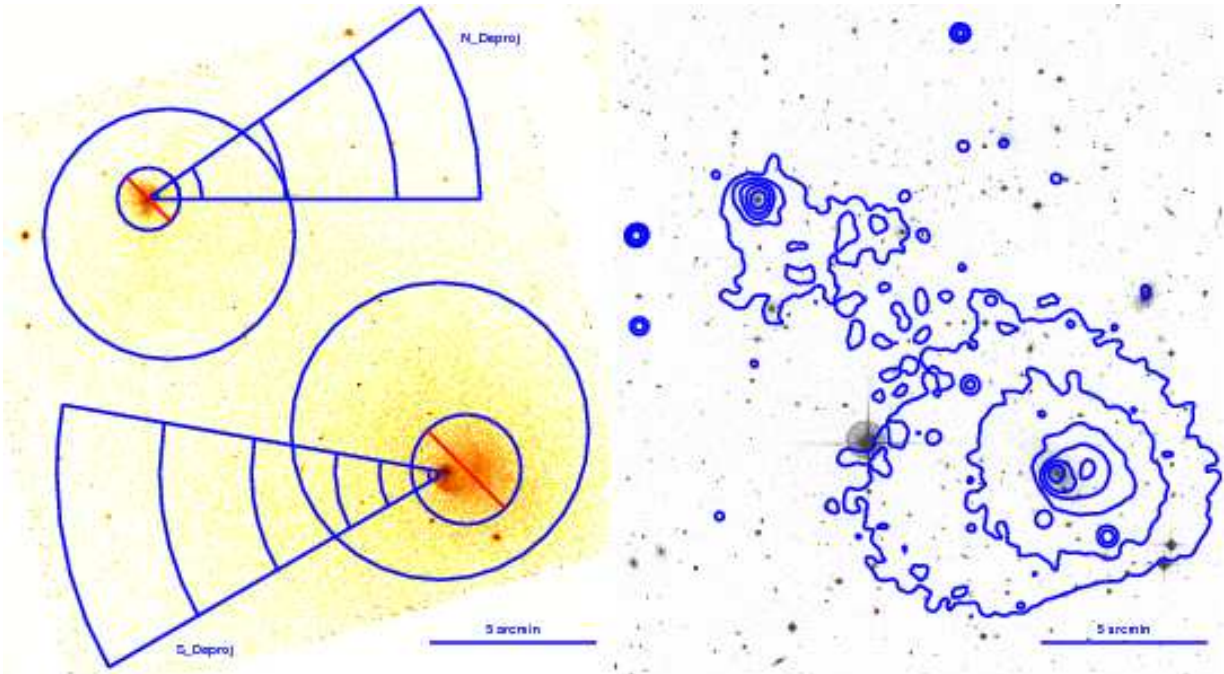}
\caption{\footnotesize{(left) Same as Figure~1 with A1644-N and A1644-S cluster regions, whose fits are in Table 2, shown as circles with the cores excluded along with outer deprojection spectral regions shown as sectors. (right) corresponding R-band optical image with X-ray surface brightness contours overlayed, showing the peaks of the X-ray emission lying on top of two bright elliptical galaxies.  A low surface brightness elongation to the SW of A1644-N extends towards the main cluster.}}
\end{figure*}

\begin{figure*}[h]
\includegraphics[scale=0.9]{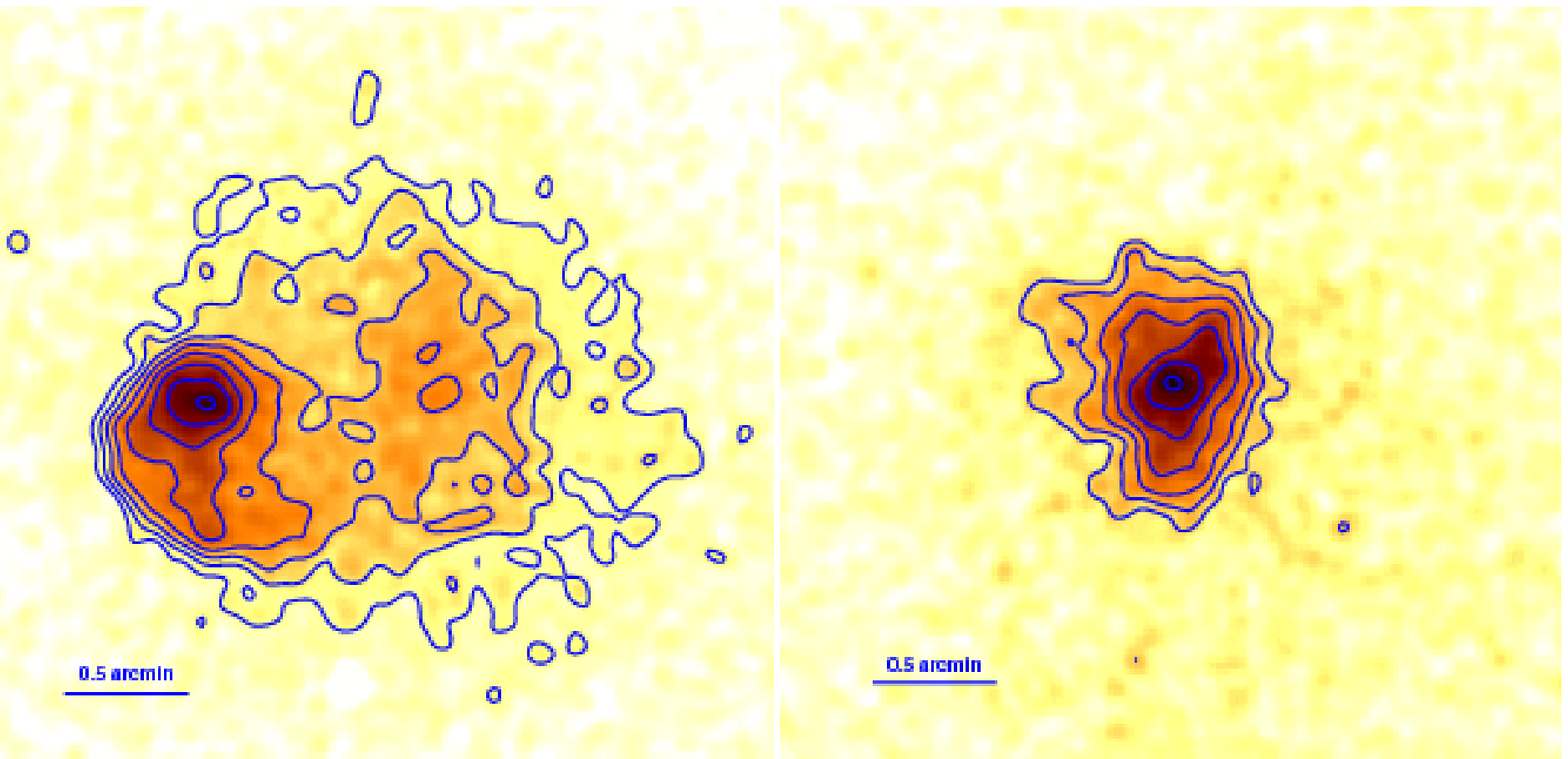}
\caption{\footnotesize{Closeup of A1644-S (left) and A1644-N (right) overlayed with isophotal contours of X-ray surface brightness. The physical image scale is 0.96 kpc/arcsec. Color gradient is increasing surface brightness from light to dark.}}
\end{figure*}

\section{PREVIOUS AND NEW OBSERVATIONS OF A1644}
We present a 72 ks \textit{Chandra ACIS-I} observation (ObsIDs 7922 and 2204) of the cluster A1644 (z=0.047), a rich binary cluster located $\sim$3 Mpc from the Shapley Super cluster (De Filippis et al. 2005), likely lying along one of the Shapley filaments (see Figure~1 and Figure~2).

The most complete sample of cluster members in A1644 contains 141 galaxies (Tustin et al. 2001). They find a lack of evidence for significant substructure in the galaxy distribution and so do not differentiate between the main and sub-cluster but instead give a mean redshift cz = $14,295\pm93$ km s$^{-1}$ and $\sigma_v$=1108 km s$^{-1}$ for the entire cluster.
Although there is disagreement in the two previous redshift studies (Dressler \& Schectman 1988; Tustin et al. 2001) as to the presence of substructure, A1644 has a clear bimodal distribution in X-rays, first observed with the \textit{Einstein} X-ray Observatory (Jones \& Forman 1984).

Subsequent \textit{XMM-Newton} observations (Reiprich et al. 2004) revealed that the two X-ray clusters (a main- and sub-cluster) are connected by a warm (4-6 keV), tenuous intracluster medium. The two X-ray peaks coincide with two of the brightest galaxies in the region. The southern, brighter X-ray peak lies on the brightest galaxy in the cluster, a cD with M$_{H}$= -25.84, while the northern, fainter X-ray peak lies on the 5th brightest galaxy, an elliptical, with M$_{H}$= -24.74 (Tustin et al. 2001). Thus we assume that the X-ray peaks and the galaxies all lie at local minima in the cluster gravitational potential. Based on the H-band magnitudes (Tustin et al. 2001), and the X-ray data, we henceforward refer to the SW cluster as the ``main'' cluster or A1644-S and the NE cluster as the ``subcluster'' or A1644-N.

\subsection{ACIS-I Data Reduction}
A1644 was observed with ACIS-I, using chips I0-3, S2, S3 in two observations (ObsIDs 7922 and 2204) with total exposure time of 72 ks. We reprocessed the level 1 event files as described in Vikhlinin et al. (2005).  The background flare rejection was done using the ratio of the 2.5-7 keV and 9.0-12 keV bands (Hickox \& Markevitch 2006) over the CCDs I0-I3 and S2, excluding bright source and cluster emission.  This resulted in a total $\sim$70 ks clean exposure, on which we performed further spectral and image processing tasks described below. Among the products created in this initial reduction are background images and exposure maps for several interesting energy bands, which incorporate the telescope aspect solution, vignetting, energy dependent effective area, and position and energy dependent detector efficiency.

Proper estimation of the X-ray background is critical for spectral analysis of the faint extended emission from clusters.  We used the blank sky datasets from the \textit{Chandra} calibration database (CALDB). The aspect solution for our observation was applied to the blank sky exposures, remapping the background events to our source coordinate frame. We then performed an overall background normalization by taking the ratio of the count rates at high energies (9-12 keV, where the telescope's effective area drops dramatically) in our source images to that in our background images. For each of our spectral extraction regions, an identical region on the detector was chosen from the blank sky image to provide a background estimate.  Associated ARFs and RMFs were created using CIAO v4.1.

\subsection{X-ray Imaging and Photometry}
From the \textit{Chandra} event file, we created an image in the 0.5-2.5 keV band in order to maximize detector sensitivity and contrast of the soft cluster emission against the detector background. We then subtracted the background image and divided the result by the corresponding exposure map. This exposure-corrected image was used for all subsequent position and imaging analyses and is shown in Figure~1 and Figure~2 (left). We detected point sources using the wavelet detection algorithm \textit{wvdecomp} outlined in Vikhlinin et al. (1998). The detected point source regions were checked by eye and then excluded from all subsequent analyses.

\subsection{Spiral Structure in the Core of A1644~-S}
\textit{Chandra's ACIS-I} field of view is large enough to image both A1644-N and A1644-S on four CCDs (I0-I3), allowing a simultaneous view of both the ICM emission connecting the two subclusters and a high resolution picture of their cores. In A1644-S, we see a spiral pattern in the core X-ray emission, defined by an edge just to the north of the brightness peak and extending over more than 180$\degr$ in azimuth (see Figure~3 left).  To examine the physical properties of the gas on either side of this prominent surface brightness edge near the core of A1644-S, we extract spectra in regions which trace this edge.  In an effort to trace the edge with circular regions, which we require for the application of local spherical symmentry in $\S$4.3, and to maximize our signal in each region, we choose the two sectors shown in Figure~4 (left).

\begin{figure*}[h]
\includegraphics[scale=0.85]{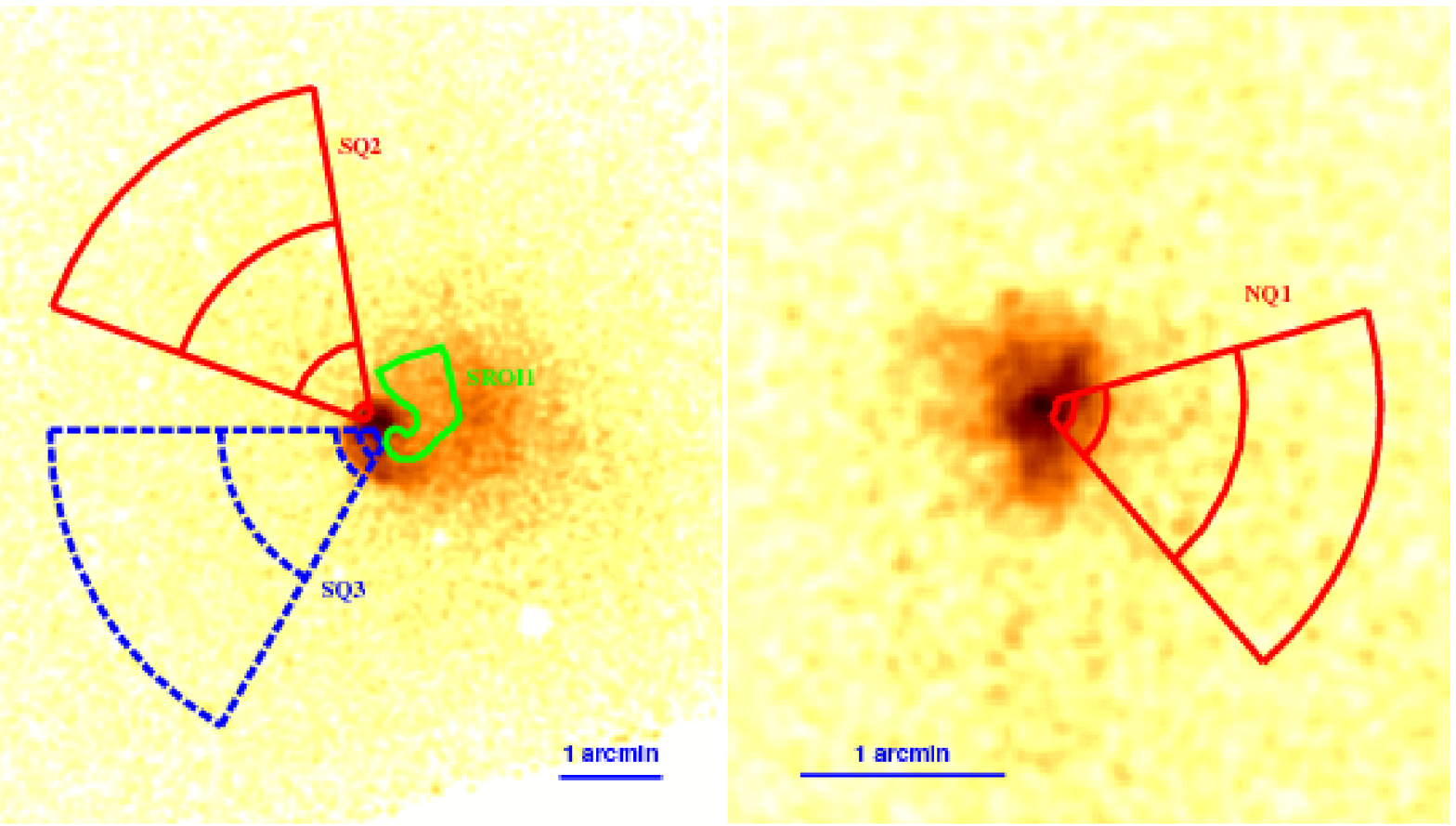}
\caption{\footnotesize{A1644-S (left) and A1644-N (right) with the spectral extraction regions overlayed.  The regions are color and pattern coded to match the points plotted in Figures 5 and 6.}}
\end{figure*}

\begin{figure*}[h]
\includegraphics[scale=0.65,angle=270]{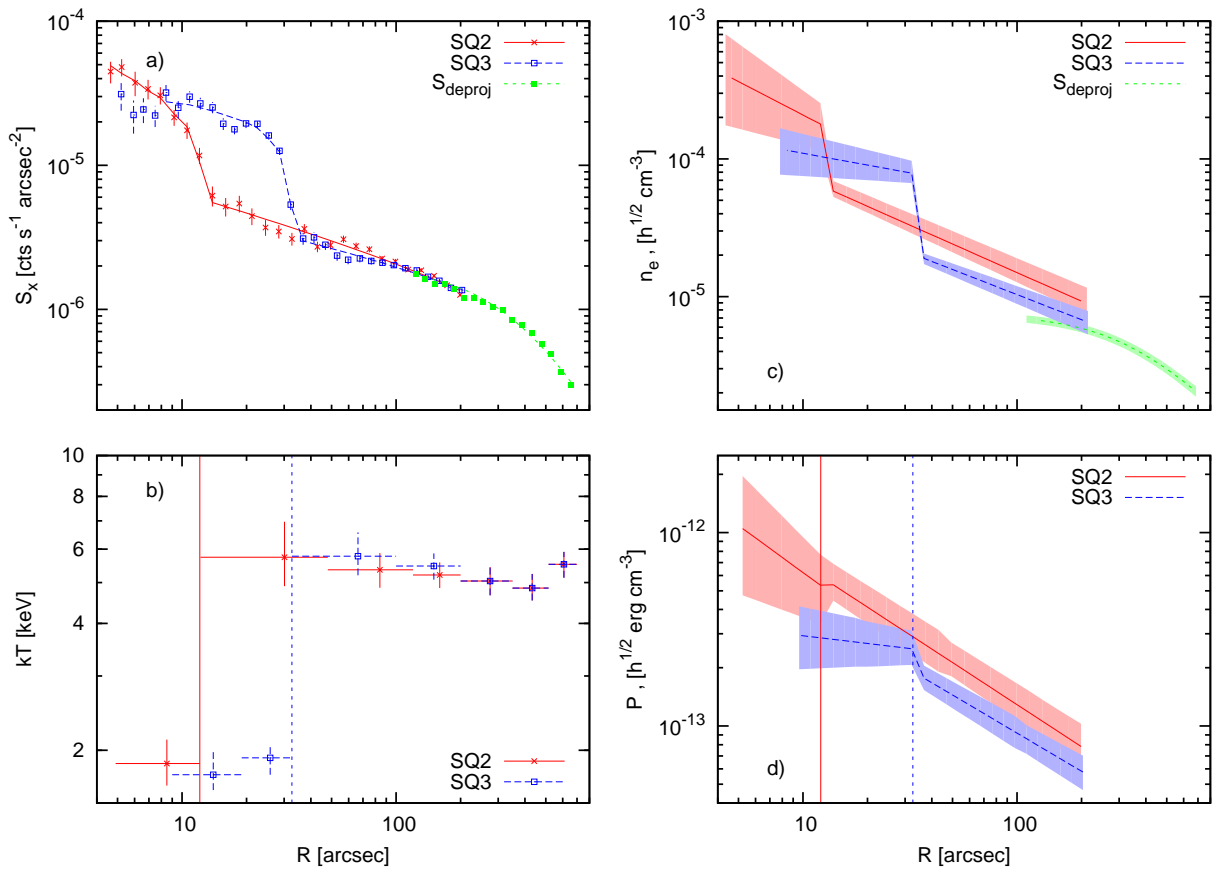}
\caption{\footnotesize{A1644-S: a) Radial surface brightness, b) deprojected temperature, c) electron density, and d) pressure profiles for the regions SQ2 (solid x's, medium (red) shading), SQ3 (open dashed squares, dark (blue) shading) in Figure~4 (left), and S$_{Deproj}$ (solid squares, light (green) shading) in Figure~2 (left). Vertical dashed lines are the radii of the edge in each sector. Spectral extraction and fitting is described in $\S$ 3.5.  Shading in c) and d) are from the 90\% confidence intervals on the model parameters.}}
\end{figure*}

\begin{figure*}[h]
\includegraphics[scale=0.65,angle=270]{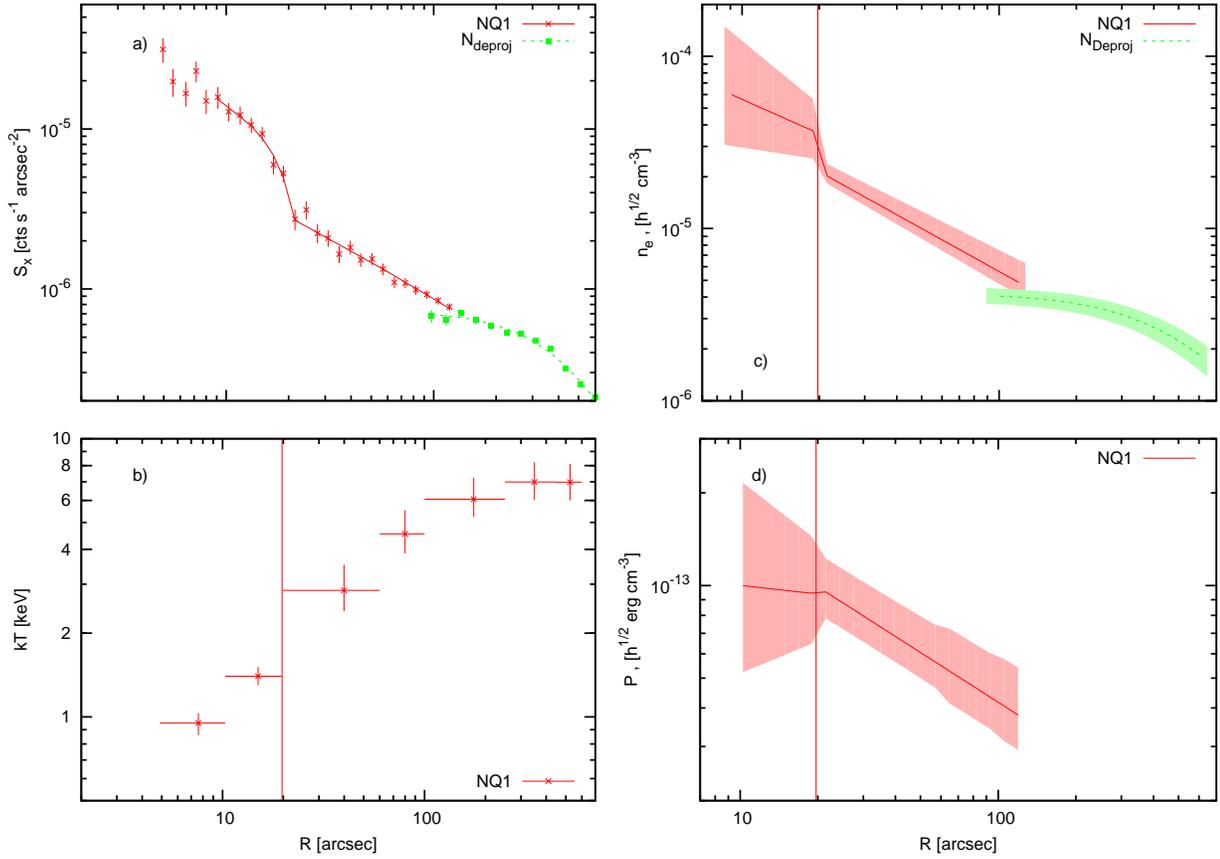}
\caption{\footnotesize{A1644-N: a) Radial surface brightness, b) deprojected temperature, c) density, and d) pressure profiles for the regions NQ1 (solid x's, medium (red) shading) in Figure~4 (right) and N$_{Deproj}$ (solid squares, light (green) shading) in Figure~2 (left). Spectral extraction and fitting is described in $\S$ 3.5  Shading in c) and d) are the 90\% confidence intervals on the model parameters.}}
\end{figure*}

\subsection{Spectral Extraction and Fitting}
Spectral extraction was performed using the CIAO\footnote{All spectral extraction and spectral analysis were performed using CIAO v4.1, the CALDB v4, and Sherpa.} tool \textit{specextract}. We defined two sectors over which to select our regions, as the projected surface brightness edge could be adequately traced by a circular region over $\sim$60 degrees in azimuth. Within each sector, spectra were extracted in circular segments at varying radii from the cluster center.  We also examine a surface brightness edge in the smaller subcluster to the north, over a sector centered on the X-ray peak where the isophotes appear compressed (see Figure~4, right).

For fitting, spectra in each region was binned with a minimum of 30 cts per bin in the inner regions (SQ2, SQ3 and NQ1), and 50 cts per bin in all other regions.  For each spectrum, we fit a thermal MEKAL model convolved with an intergalactic absorption model (xsmekal * xszwabs). The cluster's redshift was fixed at the median redshift (z=0.047) obtained from the spectroscopic sample in Tustin et al. (2001) with an absorbing hydrogen column density N$_H=5.3 \times 10^{20}$ cm$^{-2}$ (Dickey \& Lockman 1990) at the cluster position. We account for the gas lying in projection along the line of sight by using a variation of the algorithms \textit{projct} for \textit{XSpec} and \textit{deproject} for \textit{Sherpa}. The basic algorithm is to take a series of radially extracted spectra, fit a model to the outermost bin first and then move inward. For each inward step, an additional model is added to the fit, whose gas temperature (and optionally abundance) is fixed at the values obtained from the outer bins. The difference between these and our method lies in our separate derivation of the 3D gas density rather than leaving the density as a free parameter in the model fit. The density for each radial bin is calculated as described in $\S$4.3 and the ratio of the emission measure in the projected region to that in the emission region is used as a normalization constraint for each of the projected regions.

The deprojection of this system is complicated by the extended cluster emission that is preferentially distributed along the merger axis and largely fills the ACIS-I camera (see Figure~1 and Figure~2, right).  So, simply extending our spectral extraction regions (particularly SQ2) out to large radii will inaccurately predict the emission along the line of sight.  Instead, we examine sectors orthogonal to the merger axis of the cluster with radii out to the detector edge, which are likely to be more representative of the line of sight gas distribution.  We then fit a density model as described in $\S$4.3 and used this to calculate the emission measure for the outermost regions in the deprojection (labelled N$_{Deproj}$ and S$_{Deproj}$ in Figure~2). The model temperatures are listed in Table~1, plotted in Figure~5 and Figure~6 for each region in Figure~4, and are discussed below.

\begin{deluxetable}{ccccc}
\tablecaption{Radial Profiles}
\tablewidth{0pt}
\tabletypesize{\footnotesize}
\tablehead{\colhead{Region} & \colhead{R$_{in}$} & \colhead{R$_{out}$} & \colhead{kT} & \colhead{$\chi^2$ (dof)} \\
 & \colhead{($\arcsec$)} & \colhead{($\arcsec$)} & \colhead{(keV)} & }
\startdata
S$_{Deproj}$ & 515.0 & 700.0 & 5.52$^{+0.39}_{-0.39}$ & 0.97(269) \\
 & 350.0 & 515.0 & 4.85$^{+0.39}_{-0.32}$ & 0.75(263) \\
 & 200.0 & 350.0 & 5.04$^{+0.40}_{-0.38}$ & 0.76(229) \\
\tableline
SQ2 & 120.0 & 200.0 & 5.21$^{+0.37}_{-0.36}$ & 0.71(183) \\
 & 48.0 & 120.0 & 5.36$^{+0.51}_{-0.50}$ & 0.77(121) \\
 & 12.1 & 48.0 & 5.74$^{+1.23}_{-0.84}$ & 0.64(36) \\
 & 4.9 & 12.1 & 1.86$^{+0.26}_{-0.21}$ & 0.75(11) \\
\tableline
SQ3 & 100.0 & 200.0 & 5.47$^{+0.39}_{-0.39}$ & 0.68(193) \\
 & 32.6 & 100.0 & 5.77$^{+0.80}_{-0.57}$ & 0.71(92) \\
 & 19.0 & 32.6 & 1.92$^{+0.13}_{-0.17}$ & 0.95(37) \\
 & 9.0 & 19.0 & 1.75$^{+0.23}_{-0.14}$ & 1.90(23) \\
\tableline
N$_{Deproj}$ & 450.0 & 600.0 & 6.98$^{+1.14}_{-0.97}$ & 0.75(199) \\
 & 250.0 & 450.0 & 6.99$^{+1.23}_{-0.97}$ & 0.83(241) \\
 & 100.0 & 250.0 & 6.06$^{+1.18}_{-0.81}$ & 0.62(120) \\
\tableline
NQ1 & 60.0 & 100.0 & 4.55$^{+0.98}_{-0.68}$ & 1.11(51) \\
 & 19.7 & 60.0 & 2.85$^{+0.67}_{-0.45}$ & 0.60(39) \\
 & 10.3 & 19.7 & 1.40$^{+0.11}_{-0.10}$ & 0.96(12) \\
 & 4.9 & 10.3 & 0.95$^{+0.08}_{-0.09}$ & 0.77(7) \\
\enddata
\tablecomments{\footnotesize{Table~1: MEKAL model temperatures for the regions shown in Figure~2 (left) and Figure~4.  \textit{(col~1)} Spectral extraction region identifier for regions shown in Figure~4 with the exception of the deprojection regions, which are shown in Figure~2 (left). \textit{(col~2 and 3)} Inner and outer radii of sectors. \textit{(col~4)} Deprojected MEKAL model temperature. Errors on kT are the 68\% confidence interval,$\Delta\chi^2$=1, for one interesting parameter). \textit{(col~5)} Reduced $\chi^2$ for model and number of degrees of freedom.}}
\end{deluxetable}

\begin{figure*}[h]
\includegraphics[scale=0.9]{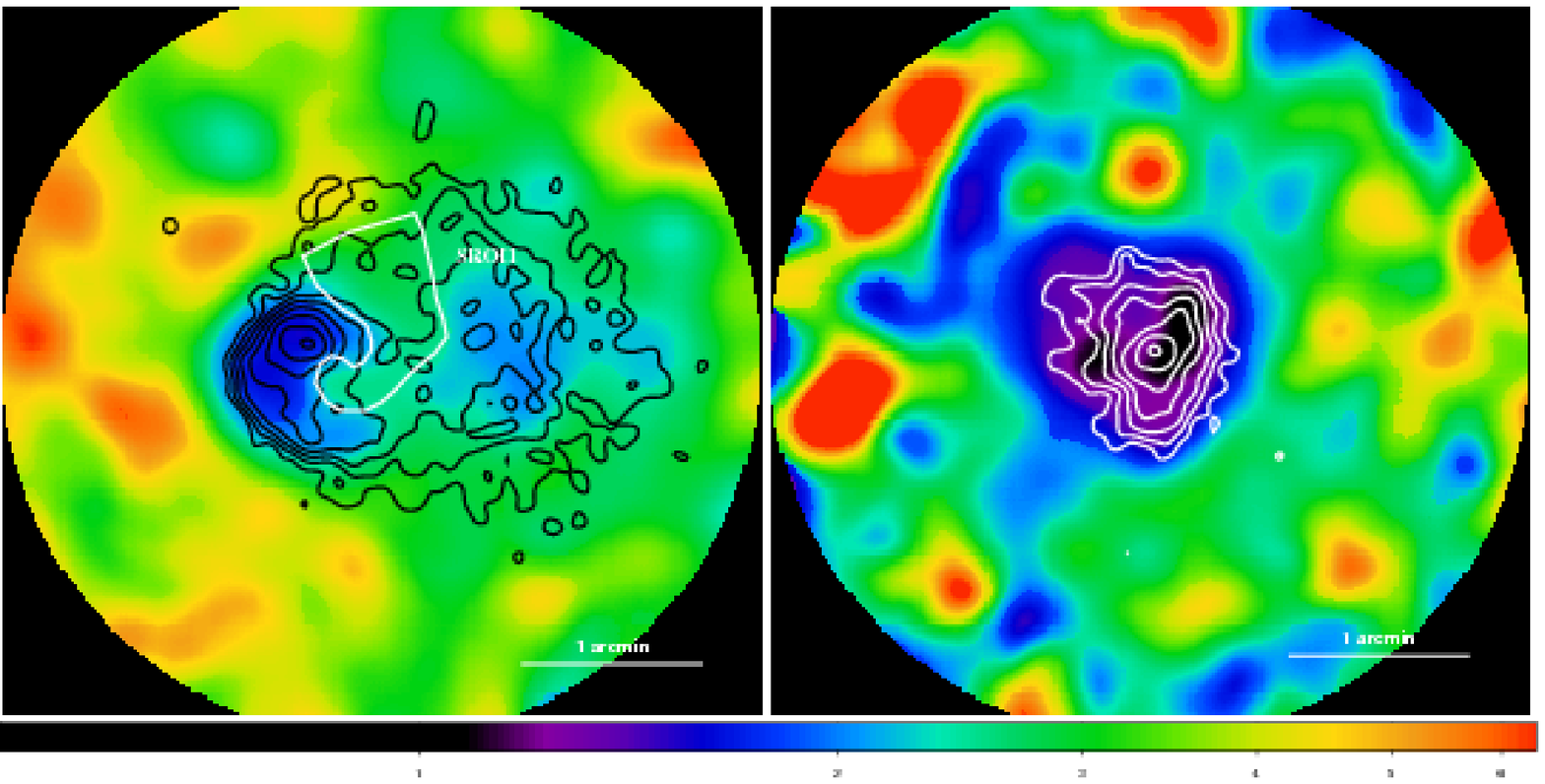}
\includegraphics[scale=0.9]{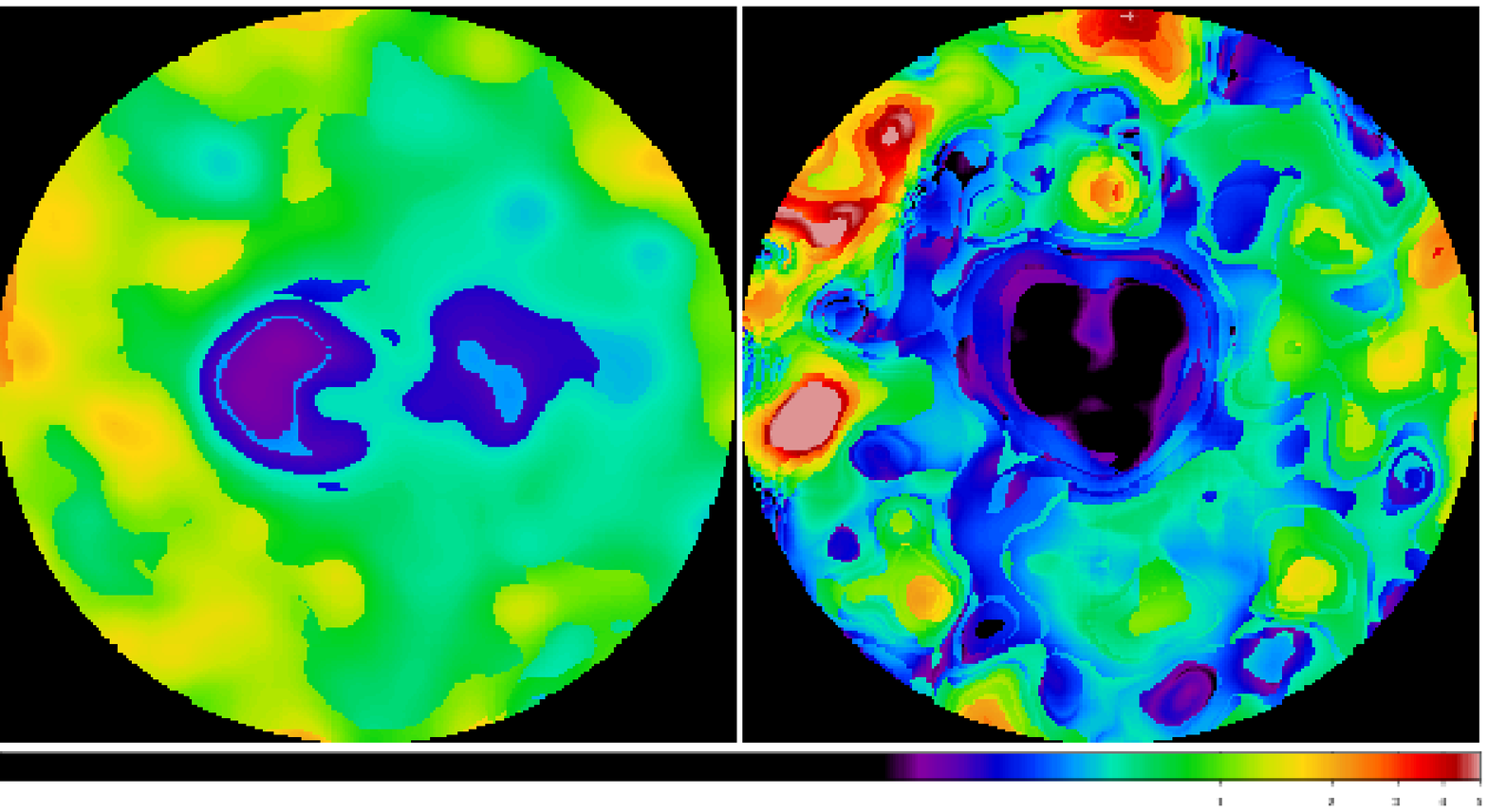}
\caption{\footnotesize{(upper) Projected temperature maps of the A1644-S (left) and A1644-N (right).  Contours are of X-ray surface brightness identical to Figure~3.  Spectral model parameters for SROI1 are shown in Table~2. Color gradient shown by color bar is temperature in keV.  (lower) Average of the hi and low 90\% confidence interval on projected temperatures.  \textit{See online version of The Astrophysical Journal for color version of this figure.}}}
\end{figure*}

\begin{figure*}[h]
\includegraphics[scale=0.45]{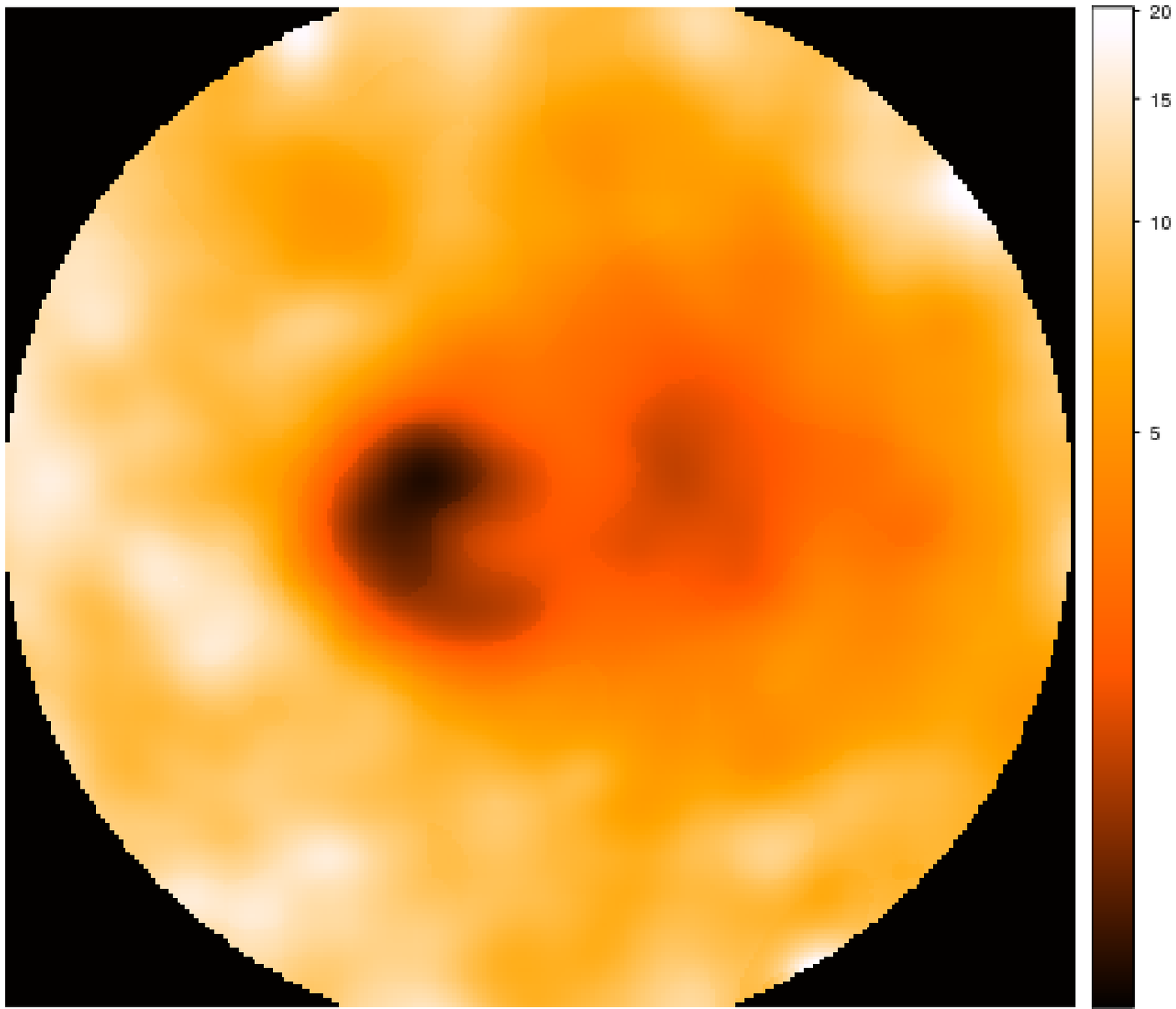}
\caption{\footnotesize{Pseudo entropy ($S=T/S_x^{1/3}$) map of A1644-S produced from the temperature map and corresponding surface brightness image (0.5-2.5 keV). We see here the finger of gas (SROI1 in Figure~7) is of higher entropy than in the core. Such streams may provide a mechanism for increasing core entropy and offsetting cooling.  \textit{See online version of The Astrophysical Journal for color version of this figure.}}}
\end{figure*}

\section{CLUSTER MEASUREMENTS}
\subsection{Main and Subcluster Data}
In Table 2, we show the best fits for several large regions around the cluster.  In the circular regions surrounding, but excluding, each of the cool cores, we find kT=$4.62^{+0.24}_{-0.16}$ keV for A1644-N and kT=$5.10\pm{0.14}$ keV for A1644-S.  These regions are shown in Figure 2 (left).  We compare the Chandra and XMM-Newton temperatures using the portion of the XMM-Newton region reported by Reiprich et al. (2004) that is covered in the smaller Chandra FOV.  Here, we obtain kT=$4.42\pm{0.05}$ keV and abundance $0.44\pm{0.03}$, as compared to their kT=$3.83\pm0.06$ keV and abundance $0.32^{+0.02}_{-0.03}$.  Abundances are reported with respect to solar using the models of Anders \& Grevesse (1989).  

Using the M$_{500}$-T$_X$ relation from Vikhlinin et al. (2009), we find M$_{500}=2.6\pm{0.4}\times10^{14} h^{-1} M_{\odot}$ for A1644-N and M$_{500}=3.1\pm{0.4}\times10^{14} h^{-1} M_{\odot}$ for A1644-S.  The latter mass is consistent with the mass derived for A1644-S from a caustic analysis in Tustin et al. (2001), where they find M$_{500}\sim4.6\pm{0.7}\times10^{14}M_{\odot}$

\begin{deluxetable}{cccc}
\centering
\tablecaption{Spectral Fits for Large Regions of Interest}
\tablewidth{0pt}
\tabletypesize{\footnotesize}
\tablehead{\colhead{Region} 	& 
	\colhead{kT} 	& 
	\colhead{Abundance} & 
	\colhead{$\chi^2$ (dof)}\\
	\colhead{} 	& 
	\colhead{(keV)} 	& 
	\colhead{} & 
	\colhead{}}
\startdata
Entire cluster & 4.42$\pm$0.05 & 0.44$\pm$0.03 & 1.66(804)  \\
A1644-N* & 4.62$^{+0.24}_{-0.16}$ & 0.44$\pm0.08$ & 1.66(403) \\
A1644-S* & 5.10$\pm$0.14 & 0.29$^{+0.05}_{-0.04}$ & 1.06(576)  \\
\tableline
SROI1 & 3.46$^{+0.20}_{-0.19}$ & 0.44 & 1.10(63) \\
\enddata
\tablecomments{\footnotesize{Table 2: MEKAL model parameters for several large regions: the portion of the cluster region described in Reiprich et al. (2004), two circular regions surrounding each cluster's X-ray peak* (shown in Figure~2, left), and SROI1 (shown in Figure~4, left).  \textit{(col~2)} Temperature,  \textit{(col~3)} abundance, and \textit{(col~4)} reduced $\chi^2$ from the MEKAL model fits.  Errors are the 68\% confidence interval on two interesting parameters, $\Delta\chi^2$=2.3, with other parameters free.  Abundance for SROI1 was fixed at the cluster mean.  *Excluding the cool cores (r=56 kpc for N and r=97 kpc for S)}}
\end{deluxetable}

\subsection{Temperature Maps}
In Figure~7 we present temperature and temperature error maps for A1644-S and A1644-N.  Temperature maps can reveal interesting structures and inhomogeneities in the cluster gas which might otherwise be overlooked. We use the method of Markevitch et al. (2000), in which a series of 6 images in non-overlapping energy bands (0.5-1.5, 1.5-2.3, 2.3-3.5, 3.5-4.5, 4.5-6.5, 6.5-9.5 keV) are created along with
corresponding exposure maps.  Each surface brightness image is then smoothed identically with a variable length scale Gaussian kernel, such that the smoothing scale at each pixel decreases roughly as the square root of the number of counts in that pixel (out to some low level of brightness, below which the smoothing scale is constant).  This method has the effect of applying a smaller smoothing scale in areas of bright cluster emission and a larger smoothing scale in the low surface brightness regions, resulting in approximately the same relative statistical accuracy across the interesting bright regions of the cluster. The surface brightness images are then used to fit a mekal model spectrum for each image pixel, binned to those same energy bands, and a chi square minimization is performed allowing only the model temperature and normalization to vary.

The temperature maps (Figure~7) for the central regions of the N and S clusters show the cool gas surrounding each cluster's X-ray peak. Their cool cores exhibit sharp temperature gradients, with projected gas temperatures going from $\sim1.5$ keV near the centers to $\sim6$ keV outside the surface brightness edge over a radial distance of $\sim20\arcsec$ ($1\arcsec\sim1$ kpc).  A small finger-shaped stream of warm, less dense (higher specific entropy) gas lies just to the west of the cool central peak in A1644-S, extending inwards to within $\sim22$ kpc of the core, labelled SROI1 in Figure~4 (left) and Figure~7 (left). In addition to cold fronts being created by core gas sloshing, simulations predict the hotter, outer cluster gas can be drawn in towards the center along such finger shaped streams (AM06; Poole et al. 2006). We discuss the implications of this further in $\S$ 5.

In A1644-N (Figure 7 right), we see a more circularly symmetric temperature structure near the core, in approximate agreement with X-ray brightness contours. There is a compression of the X-ray isophotes to the southwest and the minimum in the temperature map (corresponding to the lowest entropy gas) lies at the inner edge of this compression. This compression indicates that the core in A1644-N is moving to the west with respect to its surrounding gas. We extract spectra in a radial sector across this edge and discuss the results in $\S$5.2.

\subsection{Fitting the Surface Brightness Edges}
Figure~5a shows the surface brightness profiles of the two sectors in A1644-S (regions SQ2 and SQ3 in Figure~4 left ). We observe the edge in each sector, where the surface brightness drops abruptly at r$\sim12.1\arcsec$ ($\sim12$ kpc) in SQ2 and r$\sim32.6\arcsec$ ($\sim31$ kpc) in SQ3 from their respective centers of curvature. We select radial bins inside and outside these edges for spectral extraction, with the bin width chosen to maximize signal to noise in each region, while maintaining sufficient resolution to resolve any radial trends. The fitted gas temperatures ($\S$3.4) are shown in Table~1 and plotted in Figure~5b. They confirm that the surface brightness edge in each quadrant is indeed a cold front with the deprojected gas temperature increasing from kT=1.86$^{+0.26}_{-0.21}$ keV inside the edge to kT$=5.74^{+1.23}_{-0.84}$ keV in SQ2 and from kT$=1.92^{+0.13}_{-0.17}$ keV to kT$=5.77^{+0.83}_{-0.57}$ keV in SQ3. To quantify the magnitude of these jumps precisely, we define the temperature jump, J$_{T}$, as the ratio of the deprojected temperature measured from one side of the jump to the other:  J$_T=T_{out}/T_{in}$.  Given that the temperature on either side of the front is not uniform (in SQ3 and NQ1), we correct for the temperature gradient by modelling it with a power law on each side and obtain its value just inside and just outside the edge radius.  We present these in Table 3.

\begin{deluxetable}{ccc}
\tablewidth{0pt}
\tablecaption{Cold Front Parameters}
\tablehead{\colhead{Region ID} &  \colhead{$J_T$} & \colhead{$J_{n_e}$}}
\startdata
SQ2 & $3.09^{+1.13}_{-0.78}$ & 2.62$^{+1.47}_{-1.19}$ \\
SQ3 & $3.01^{+0.74}_{-0.47}$ & 3.79$^{+1.32}_{-0.83}$ \\
NQ1 & 1.88$^{+0.66}_{-0.50}$ & 1.98$^{+0.89}_{-0.79}$ \\
\enddata
\tablecomments{\footnotesize{Table 3: Jumps in temperature and density across the regions in Table 1.  Ratio of temperature  \textit{(col~2)} and density \textit{(col~3)} inside and outside the surface brightness edges as described in $\S$4.3.  Errors on the density jump are the 90\% confidence interval.}}
\end{deluxetable}

Since the plasma X-ray emissivity is proportional to the square of the gas density (and only weakly dependent on the temperature for our energy band), the surface brightness distribution provides a direct estimate of the gas density. A discontinuity in the slope of the surface brightness profile at the edge therefore implies a discontinuity, or a jump, in the gas density.  We model the density within our spherical sectors encompassing the edge with a power law function immediately interior to the edge (as in Markevitch et al. 2000 and many later works). Outside the edge in A1644-S and A1644-N, a second power law profile gives a good fit, whereas for the larger deprojection regions a $\beta$ model is used. The density models are:

\begin{eqnarray*}
n_e(r)&=&n_1 J_{n_e}(\frac{r}{r_j})^{\gamma_1}, \\ && \text{ for } r<r_j \text{ in all regions} \\
&=&n_2N(\frac{r}{r_j})^{\gamma_2}, \\ && \text{ for } r>r_j \text{in SQ2, SQ3 and NQ1}\\
&=&n_2S[1+(\frac{r}{a})^{^2}]^{-\frac{3\beta}{2}}, \\ && \text{ for } r>r_j \text{ in S$_{Deproj}$ and N$_{Deproj}$}
\end{eqnarray*}
where $r_j$ is the radius of the edge \footnote{In general, the centers of the inner and outer models may differ, however they are assumed identical}, $\gamma_1$ is the power law slope interior to each edge, $\gamma_2$ is the power law slope outside the edge, $a$ is the scale radius of the $\beta$ model, $n_1$, $n_2$ and $n_3$ combine the respective model normalizations with the emission measure normalization for each model in their respective regions, and J$_{n_e}$ is the magnitude of the density jump. When projected, this density model produces an X-ray brightness edge with the characteristic profile observed at cold fronts.  Inside the edge we limit the fit to the immediate vicinity of the edge, as the profile near the brightness peak is like to have a more complex shape.

To fit the density change across the jump, this radial density profile is squared and integrated along the line of sight, multiplied by a factor close to 1 to account for the weak dependence of the X-ray emissivity on temperature, and then fit to the surface brightness distribution. The resulting density fits are shown in Figure~5c and Figure~6c along with the model errors. We then compute the internal pressure of the gas simply as $P=n_e T$ which is plotted in Figure~5d and Figure~6d.  Figure 5 and Table 3 show that the edge in A1644-S is a very prominent cold front, while that in A1644-N is marginally consistent with a smooth gas profile.

\begin{figure*}[h]
\includegraphics[scale=0.65]{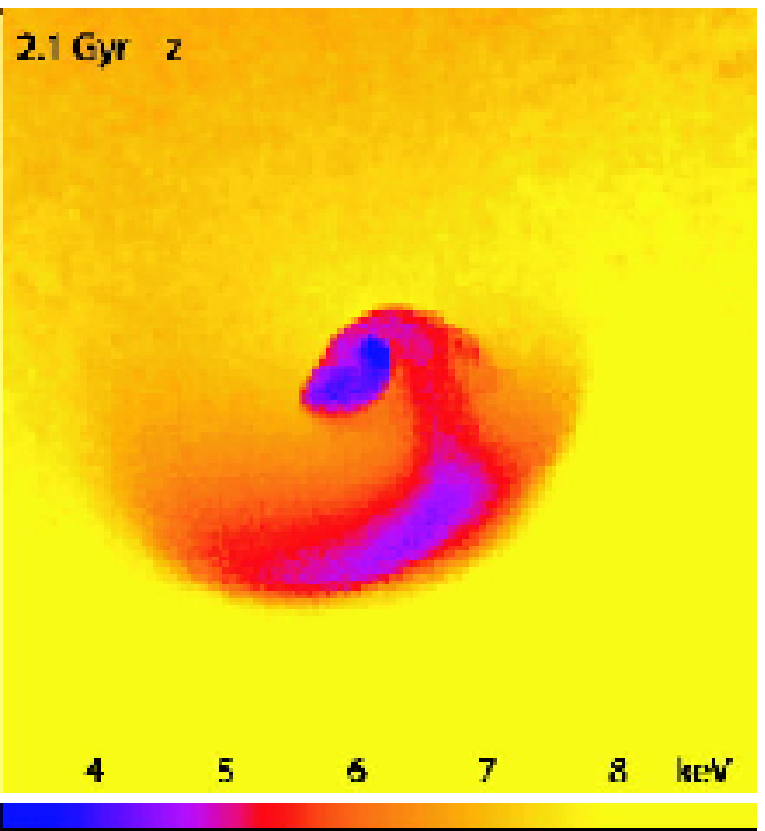}
\includegraphics[scale=0.65]{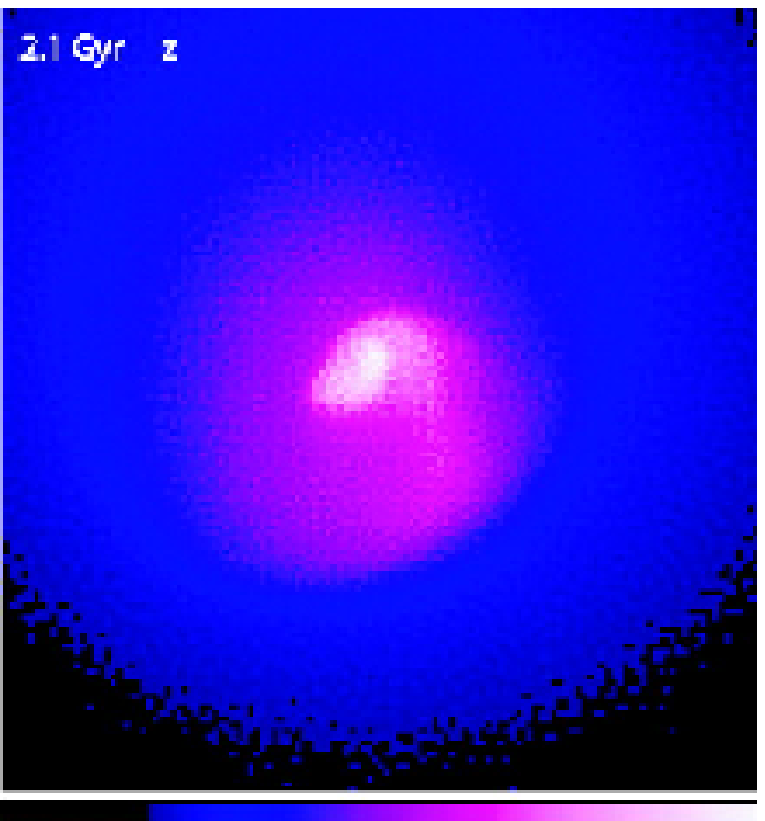}
\includegraphics[scale=0.45]{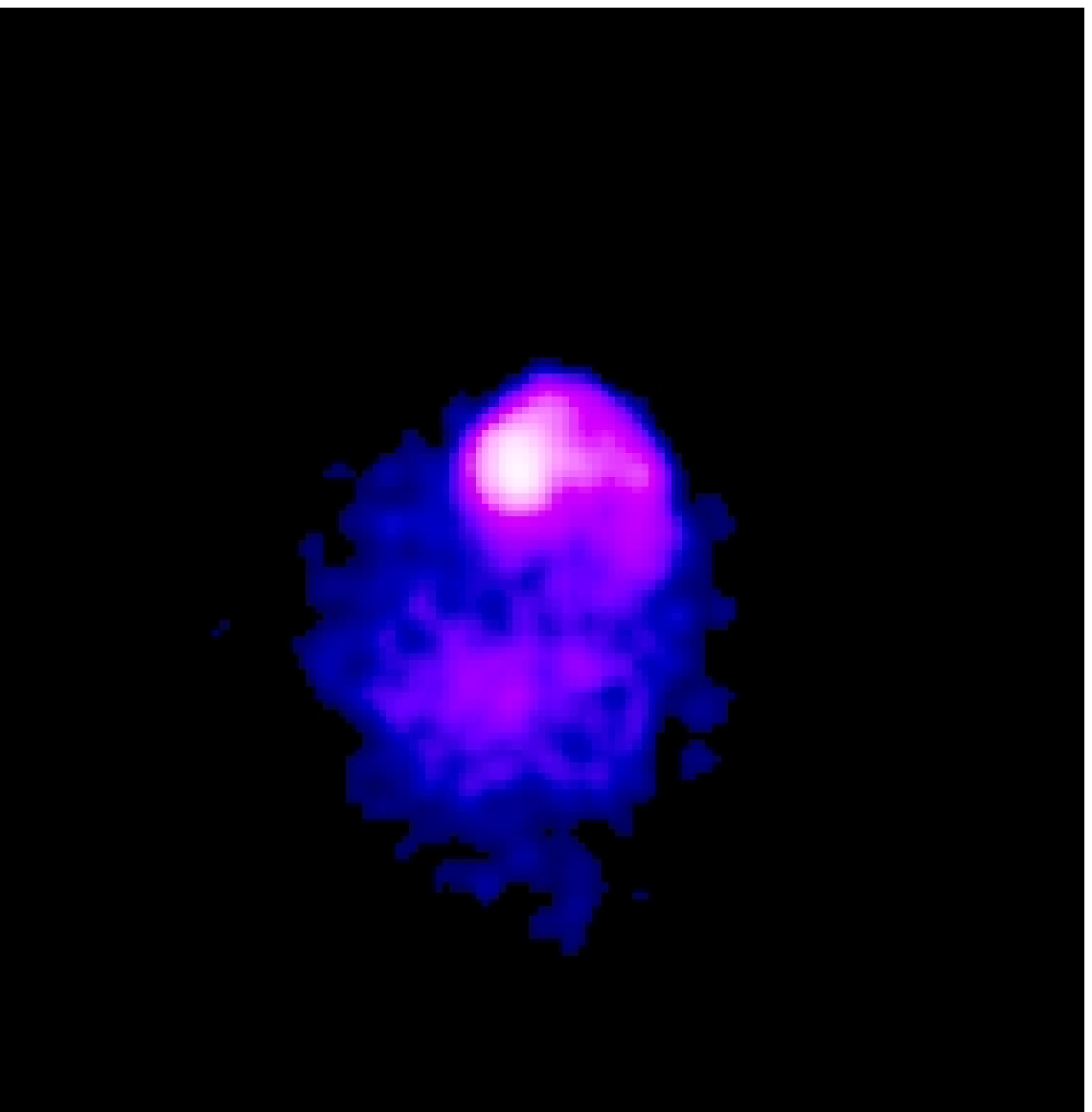}
\caption{\footnotesize{Projected temperature (left) and X-ray surface brightness (center) images of the R=5 merger of two clusters reproduced from AM06 by permission of the AAS. (right) A smoothed 0.5-2.5 keV image of south cluster of Abell 1644, oriented to match the simulation frames (90$\%$ clockwise rotation then Y-axis mirror reflection). The physical scales on each of these images are approximately 250 kpc on a side.}}
\end{figure*}

\section{DISCUSSION}
\subsection{Comparison With Simulations}
To assist in interpreting the gas motion which produced the spiral morphology, we reproduce a snapshot from the simulations of AM06 in Figure~9 (also see Figure 7 in AM06).  Although this simulation is for an R=5 mass ratio with a gasless subcluster, other simulations in AM06 show a qualitatively similar structure on the small scales we are comparing here ($r<100$ kpc). So our conclusions may be
  equally well drawn from cases with initial conditions more appropriate to A1644 (e.g., the $R=2$, subcluster-with-gas case shown in Fig.\ 14 of AM06; note that on large scales, A1644 does show the disturbed morphology consistent with a merger with a subcluster with gas).  Our choice here was made as it affords the highest-resolution morphological comparison to already published data.

At the 1.6 Gyr snapshot, the perturber has already passed above the main cluster, from the top right towards the bottom left in the simulation frames. We find the 2.1 Gyr snapshot in the AM06 simulations (700 Myr after pericentric passage of the subcluster) to be the most similar image to the present state of A1644 (see Figure~9 where we show the simulations and our image side by side). The cool core gas, which had previously been in equilibrium with the cluster potential, was disturbed by the passage of A1644-N and began sloshing around the minimum of A1644-S's gravitational potential.  The simulations predict that relatively long lived ($1$ Gyr) cold fronts can be produced by each passage of the gas core about the potential minimum, with their centers of curvature at approximately the location of the potential minimum of the cluster. Given that we see only one such cold front to the east of A1644-S, it appears that we are viewing an early stage of gas sloshing.

\subsection{Merger History}
The spiral morphology of core gas sloshing when viewed at or near edge-on provides a unique method to probe the merger history of the system. Simulations (AM06) show that following the spiral structure from the outside inwards indicates the direction of the infall of the perturbing object. In A1644, the direction of the spiral inflow is clockwise (angular momentum vector is into the page) and so the perturber must also have a clockwise orbit. To produce the observed spiral structure, the perturber most likely passed the main cluster on the eastern side heading in a northerly direction. This is the same trajectory of A1644-N as indicated by Reiprich et al. (2004). 

Adopting this model, we explore its consequences and test whether it agrees with other observational evidence. We note that the projected separation of the two clusters is $\sim12.4\arcmin$ ($\sim700$ kpc).  The absence of obvious shock fronts suggests M$\leq1.0$, so if we assume the velocity of the subcluster relative to the main cluster is $\sim900$ km s$^{-1}$ (the sound speed of the 5 keV cluster gas), we may estimate a timescale for the pair to arrive at their current projected separation.

Travelling at this average speed, the clusters will have obtained this separation $\sim 760$ / cos \textit{i} Myr after their pericentric passage, where \textit{i} is the angle of inclination of the orbit to the plane of the sky. Returning to the simulations of AM06, we see that the time from the pericentric crossing of the two cores to when the first cold front morphology is similar to Abell 1644 is $\sim 700$ Myr (the 1.4 to 2.1 Gyr snapshots), in rough agreement with the timescale for the clusters' current separation. We may also compare this to the R=2 and R=5 (subcluster with gas) in AM06 (their Figure 9 and Figure 14) and find that a similar morphology is produced after $\sim$500 and $\sim$600 Myr, again in good qualitative agreement with our calculation.  The isophotes of X-ray surface brightness in the northern subcluster, along with its projected temperature structure suggest that its core is currently moving west-southwesterly, back towards A1644-S, although they may also indicate merger induced sloshing in that core as well.

\subsection{Core Heating}
The X-ray data indicate that the passage of a subcluster has displaced the cool core of the main cluster from the gravitational potential minimum. As it oscillates, the newly acquired potentital energy will dissipate into heat.  It is interesting to compare the rate of such heating with the rate of radiative cooling in the core.

To find the energy necessary to displace its gas core from the potential minimum, we require the gravitational potential of A1644-S.  For an estimate of the gravitational potential, we use the mass-temperature relation from Vikhlinin et al. (2009) and assume an NFW (Navarro, J. et al. 1995) form of the mass profile.  Even though the cluster is disturbed, this is sufficient for our order of magnitude estimate.  Obviously, our result is roughly dependent on shape, however we note that cold fronts in the cluster indicate the presence of a potential cusp (as in a NFW profile) as opposed to a shallow cored ($\beta$-model) distribution (AM06).  

The potential energy difference between the current location of the core (the X-ray brightness peak) and the potential minimum (the center of the cD) galaxy is $U\sim1.3\times10^{56}$ erg (see Appendix A for detailed calculation). Assuming a fraction \textit{f} of that energy is converted to thermal energy in the gas during the time between pericentric crossings of the two cores ($\sim2.6$ Gyr in the AM06 simulations), we may express the resulting heating rate as P$_{slosh}=f\times1.5\times 10^{39}$ erg s$^{-1}$ and compare that to the observed X-ray luminosity in the core of A1644-S (L$_{core}=2.4\times10^{42}$ erg s$^{-1}$). Given the strong dependence of our result on the gas core radius, which is seen in projection, this is a lower limit to the available energy in the sloshing gas.  Additionally, the separation between the gas core and the potential minimum is also certainly a minimum estimate, as we are not only seeing it in projection but we are also assuming that the current separation is the maximum separation.  

Core gas sloshing may reheat the core also by drawing in higher entropy outer cluster gas along streams towards the center (MV07). A look at the entropy map (Figure~8) shows that the stream in A1644-S does contain a higher entropy plasma than that found in the core, suggesting an origin outside the core.  Detailed hydrodynamic simulations are needed to estimate the magnitude of this effect and will be addressed in a future paper (Zuhone et al. in prep).

\section{SUMMARY}
We present an analysis of the 70 ks \textit{Chandra ACIS-I} observation of the double cluster Abell 1644. Based on the X-ray and available optical data, we find compelling evidence that the primary cluster has undergone an interaction with its northern companion and that the interaction has disturbed the gas around each cluster's core. An examination of the X-ray surface brightness edge in the southern, main cluster finds that the edge is one continuous cold front with a spiral morphology arising from the core gas sloshing as seen in the hydro simulations of AM06 and across which the temperature jumps by a factor of $\sim$3.

We find that the gravitational potential energy transferred to the main cluster's gas core by the perturbing subcluster is insufficient to offset the X-ray luminosity.  Still, the spiral structure in A1644-S presents an interesting example of how merger-induced sloshing may draw higher entropy (i.e. low-density and high T) outer gas inside the cooling core, possibly supplying energy to partially offset cooling.

\section{Acknowledgements}
REJ would like to thank Alexey Vikhlinin, John Zuhone, Scott Randall, Bill Joye, Liz Galle and Nina Bonaventura for useful discussions and help in the \textit{Chandra} data processing. REJ acknowledges support for this work from a Dartmouth Graduate Fellowship along with a Smithsonian Astrophysical Observatory Predoctoral Fellowship provided by NASA through \textit{Chandra} Award Number NNX08AD89G, CXO GO6-7094X and GO6-7095X issued by the CXC, which is operated by the Smithsonian Astrophysical Observatory under contract NAS8-03060. This research has made extensive use of SAOImage DS9, in addition to software provided by the CXC in the application packages CIAO, ChIPS, and Sherpa.

\bibliography{paper}
\noindent
\\
Anders E. \& Grevesse N. 1989, \gca, 53, 197\\
AM06, Ascasibar, Y. \& Markevitch, M. 2006, \apj, 650, 102 \\
Churazov, E. et al. 2000, \aap, 356, 788\\
Churazov, E. et al. 2003, \apj, 590, 223\\
De Filippis, E. et al. 2005,\aap, 444, 387\\
Dickey, J. \& Lockman, F. 1990, \araa, 28, 215\\
Dressler, A. \& Schectman, S.A. 1988, \aj, 95, 985\\
Fabian, A. et al. 2000, \mnras, 318, L65\\
Hickox, R. \& Markevitch, M. 2006, \apj, 645, 95 \\
Jones, C. \& Forman, W. 1984, \apj, 276, 38\\
Jones, C. \& Forman, W. 1999, \apj, 511, 65\\
Lagana, T. et al. 2009, arXiv:0911.3785, accepted for publication in \aap\\
Markevitch, M. et al. 2000, \apj, 541, 542\\
Markevitch, M. et al. 2001, \apj, 562, L153 \\
Markevitch, M. et al. 2002, \apj, 567, 27\\
Markevitch, M., Vikhlinin, A., \& Forman, W.R. 2003, in ASP Conf. Ser. 301, Matter and Energy in Clusters of Galaxies, ed. S. Bowyer \& C.-Y. Hwang. (San Francisco, CA: ASP), 37\\
MV07, Markevitch, M. \& Vikhlinin, A. 2007, \physrep, 443, 1\\
Navarro, J. et al. 1995, \mnras, 275, 720\\
Nulsen, P. et al. 2002, \apj, 568, 163\\
Nulsen, P. et al. 2005, \apj, 628, 629\\
McNamara, B. \& Nulsen, P. 2007, \araa, 45, 117 \\
Peres, C.B. et al. 1998, \mnras, 298, 416\\
Poole, G. et al. 2006, \mnras, 373, 881\\
Reiprich, T. et al. 2004, \apj, 608, 179\\
Tittley, E. \& Henriksen M. 2005, \apj, 618, 227 \\
Tustin, A. et al. 2001, \aj, 122, 1289\\
Vikhlinin, A. et al. 1998, \apj, 502, 558\\
Vikhlinin, A. et al. 2001, \apj, 551, 160\\
Vikhlinin, A. et al. 2005, \apj, 628, 655\\
Vikhlinin, A. et al. 2006, \apj, 640, 691\\
Vikhlinin, A. et al. 2009, \apj, 692, 1033\\

\appendix
%\begin{appendix}
\section{APPENDIX A}

Here, we estimate the energy required to displace the core gas in A1644-S to its current (projected) position. The general expression for the gravitational potential energy of the gas core is:
\begin{center}\begin{equation}
U={G M_T(<R) m_g R^{-1}}
\end{equation}\end{center}
where $R=1.5\arcsec$ (1.44 kpc) is the distance the core gas has been displaced from the potential minimum, assuming it is a solid body for this order of magnitude estimate, M$_T(<$R) is the total cluster mass within the displaced radius, and m$_g$ is the mass of the displaced core.  

To find M$_T(<$R), we fit an NFW profile using the concentration parameter c$_{500}$=3 for a 5 keV cluster (Vikhlinin et al. 2006).  We then normalize the resultant mass profile at r$_{500}$ to the M-T relation in Vikhlinin et al. (2009) to obtain M$_T(<$R)=6.7 $\times 10^9 M_{\odot}$. 

For $m_g$, the displaced core gas mass, we integrate the gas density profile obtained from the X-ray surface brightness distribution in $\S$4.2 from the center out to the front radius r$_j$ in SQ3 ($r_j\sim33\arcsec$ or 31 kpc). From the density profile in SQ3 we obtain:
\begin{center}\begin{equation}
m_g=4 \pi k \int_{0}^{r_j} \frac{m_p}{1.17} n_e(r) r^2 dr
\end{equation}\end{center}
where $n_e(r)$ is the same as in $\S$4.3.  

This results in a radial dependence for the core gas mass of:
$m_g(r) \varpropto r_1^{3+\gamma_1}$.
The core gas mass displaced within r$_j$ is then: $m_g \sim 3.1 \times 10^8 M_{\odot}$. 
Substituting $M_T$ and $m_g$ into A1, we obtain $U \sim 1.3 \times 10^{56}$ erg which, averaged over the time between pericentric core passages (2.6 Gyr) is $\sim 1.5 \times 10^{39}$ erg s$^{-1}$.
%\end{appendix}

\end{document}